# Multimodal *Operando* X-ray Mechanistic Studies of a Bimetallic Oxide Electrocatalyst in Alkaline Media

*Jason J. Huang,[1] Yao Yang,[2†] Daniel Weinstock,[1] Colin R. Bundschu,[3] Jacob P. C. Ruff,[4] Tomás A. Arias,[3] Héctor D. Abruña,[2*] Andrej Singer[1*]*

1. Cornell University, Dept. of Materials Science & Engineering
2. Cornell University, Dept. of Chemistry & Chemical Biology
3. Cornell University, Dept. of Physics
4. Cornell High Energy Synchrotron Source

†. Present Address: Miller Institute, Berkeley Chemistry

*. Corresponding author

**Abstract**

Furthering the understanding of the catalytic mechanisms in the oxygen reduction reaction (ORR) is critical to advancing and enabling fuel cell technology. In this work, we use multimodal *operando* synchrotron X-ray diffraction (XRD) and resonant elastic X-ray scattering (REXS) to investigate the interplay between the structure and oxidation state of a Co-Mn spinel oxide electrocatalyst, which has previously shown ORR activity that rivals Pt in alkaline fuel cells. During cyclic voltammetry, the electrocatalyst exhibited a reversible and rapid increase in tensile strain at low potentials, suggesting robust structural reversibility and stability of Co-Mn oxide electrocatalysts during normal fuel cell operating conditions. At low potential holds, exploring the limit of structural stability, an irreversible tetragonal-to-cubic phase transition was observed, which may be correlated to reduction in both Co and Mn valence states. Meanwhile, joint density-functional theory (JDFT) calculations provide insight into how reactive adsorbates induce strain in spinel oxide nanoparticles. Through this work, strain and oxidation state changes that are possible sources of degradation during the ORR in Co-Mn oxide electrocatalysts are uncovered, and the unique capabilities of combining structural and chemical characterization of electrocatalysts in multimodal *operando* X-ray studies are demonstrated.

## Introduction

Ever-growing environmental concerns and global energy demand have pushed the development of fuel cells, which generate energy by electrochemically converting hydrogen and oxygen to water. Proton exchange membrane fuel cells (PEMFCs) are widely recognized as a key technology for hydrogen-based energy conversion.[1,2] Nevertheless, PEMFCs rely on expensive Pt-based electrocatalysts to facilitate the oxygen reduction reaction (ORR).[3–6] Alkaline fuel cells (AFCs) have emerged as a promising alternative since they enable the use non-precious metal electrocatalysts.[5,6] Among the alternative electrocatalysts for the ORR, 3d transition metal oxides have drawn interest due to their high activity, long durability, and low cost.[7]

A class of Co and Mn based spinels have been of particular interest since they exhibit higher activity than Pt in alkaline media. The reason for high activity is the ability of the material to use Co and Mn co-active sites to synergistically catalyze the ORR.[8,9] Previous *ex situ* scanning transmission electron microscopy (STEM) studies also revealed electrochemical activity decay caused by particle aggregation and surface area loss.[8] While *ex situ* studies of Co-Mn spinels have revealed key information, it is very unlikely that *ex situ* characterization adequately elucidates the processes that occur under real-time, dynamic electrochemical conditions that are furthermore, far from equilibrium. Therefore, *in situ* and *operando* techniques are key to understanding/unravelling the ORR mechanism and degradation pathways in Co-Mn spinel oxides.

Previously, *in situ* X-ray absorption spectroscopy (XAS) was used to identify the active sites in the bimetallic Co-Mn spinel electrocatalyst with demonstrated high ORR activity.[10] Co and Mn oxidation states were found to change simultaneously under both constant applied potential and cyclic voltammetry. This coordinated change in oxidation state indicated that the $Co^{+2/3}$ and $Mn^{+2/3/4}$ redox couples serve as co-active sites to synergistically catalyze the ORR.[10] While XAS was key to revealing the synergistic mechanism with Co and Mn active sites, questions remain about how the nanoparticle structure and morphology affect the ORR activity and nanoparticle stability. Connecting changes in the crystalline structure and oxidation state information is critical to deepening the understanding of oxide spinel ORR electrocatalysts and to designing more efficient and durable low-cost fuel cell electrocatalysts.

*Operando* X-ray diffraction (XRD) enables real-time observation of crystalline structure changes in an operating electrochemical cell. The position and shape of Bragg reflections as a function of applied potential can provide dynamic information about lattice constants, particle size, and structural disorder. The brilliance of 3rd and 4th generation synchrotron sources has enabled high-quality and high temporal resolution *operando* diffraction measurements. *Operando* XRD has been used to reveal dynamic strain, structural disorder, and surface oxidation in noble metal (Pt, Pd) fuel cell electrocatalysts.[11–13] In this work, we use *operando* XRD at the Center for High-Energy X-ray Sciences (CHEXS) at CHESS to study structural dynamics in Co-Mn spinel electrocatalysts.

In addition to *operando* XRD, we use *operando* resonant elastic X-ray scattering (REXS) to study Co-Mn spinel electrocatalysts. *Operando* REXS combines scattering and spectroscopy. During REXS, the intensity of a Bragg reflection is recorded while the X-ray energy is scanned across an absorption edge of the atomic species being studied. REXS generates spectra equivalent to XAS for large, single-phase crystals.[14] When multiple structural phases are present, REXS can provide additional information. XAS identifies chemical species and oxidation state information by observing at which X-ray energies bound electrons can be promoted to a higher energy state. The resulting absorption signal averages over resonant atoms that the X-ray beam passes through, independent of the local structure around the absorbing atoms. *Operando* REXS spectra are collected on a Bragg reflection, which allows separating spectra from different structural phases when the Bragg reflections are separated in reciprocal space. *Operando* REXS, therefore, enables observation of oxidation state information of selected crystalline phases by leveraging the energy-dependent squared magnitude of the structure factor measured as diffraction intensity in Bragg scattering.[14–16] The technique was previously used to discriminate how the nickel oxidation states in co-existing structural phases change during a phase transition in a sodium-ion battery layered cathode during charge.[17] Here, we use a multimodal approach, *operando* XRD and REXS, to observe structural strain and structure-specific Co and Mn oxidation state changes in $MnCo_2O_4$ spinel nanoparticles during both constant and dynamic potential conditions in a functioning electrochemical cell.

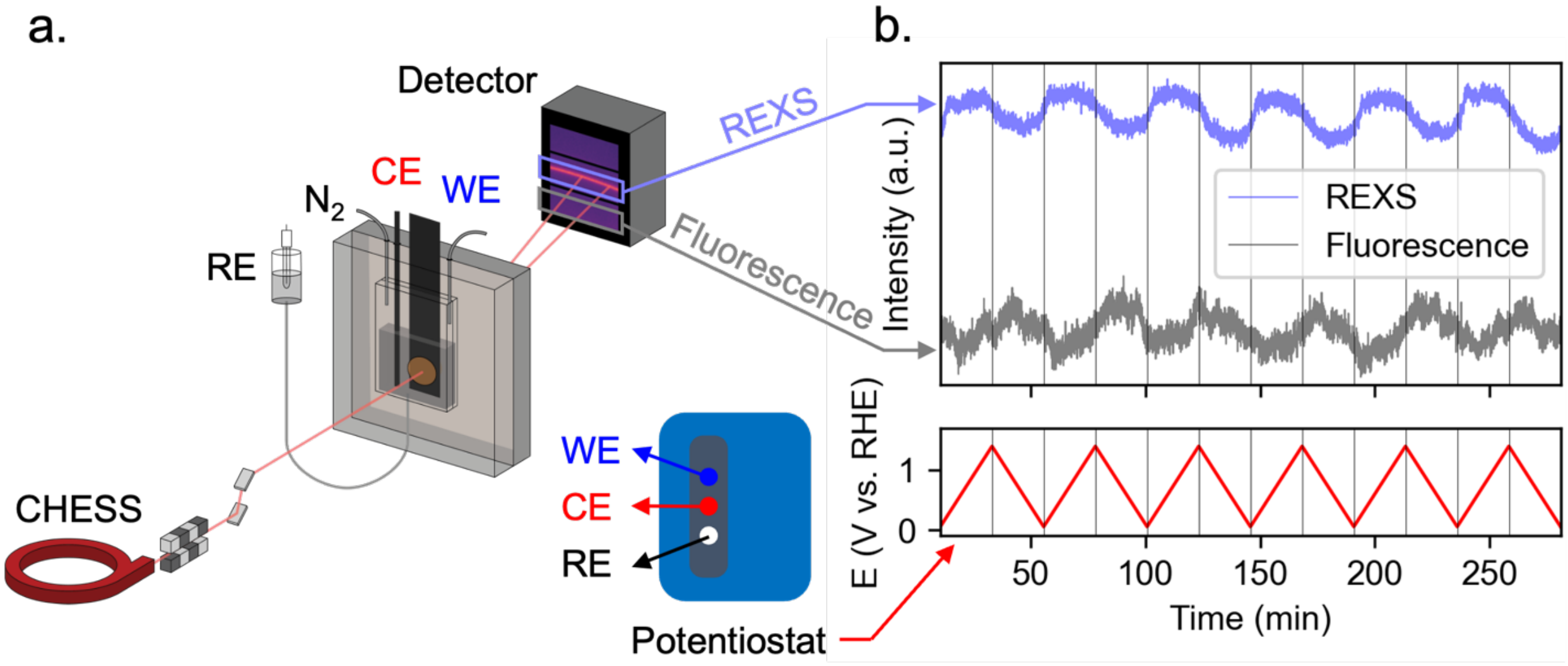

*Figure 1. Operando X-ray diffraction and Resonant Elastic X-ray Scattering. (a.) Schematic of the operando electrochemical cell and X-ray diffraction experiment. (b). $MnCo_2O_4$ 211 resonant elastic X-ray scattering (REXS) peak intensity and background fluorescence intensity at the Co K-edge (7722 eV) during cyclic voltammetry (shown in blue and grey, respectively). Applied potential during cyclic voltammetry from 0.05 to 1.4 V (shown in red).*

## Results

To understand how the Co-Mn oxide structure changes under dynamic potential conditions, cyclic voltammetry was conducted on an *operando* X-ray electrochemical cell,[10] depicted in Figure 1a. The 3-electrode cell used a working electrode composed of $MnCo_2O_4$ crystalline nanoparticles (particle size of order 30 nm) dispersed on carbon paper, a graphite rod as a counter electrode, and $Ag^+$/AgCl as a reference electrode connected via a salt bridge (more info about the cell is available in the Supporting Information). Diffraction patterns of the 211 tetragonal spinel peak were collected continuously during cyclic voltammetry with potential limits of 0.2 to 1.4 V vs. the reversible hydrogen electrode (RHE) and 0.05 to 1.4 V vs. RHE (all potential values are in the RHE scale). The upper and lower potential limits of 1.4 and 0.2 V were selected to simulate the normal fuel cell operating conditions and the events of large polarizations of oxygen cathodes during startup and shutdown procedures in realistic automotive applications.[18] In addition, the lowest potential limit of 0.05 V was chosen to induce an extreme ORR condition to examine the limit of structural stability while avoiding the hydrogen evolution reaction, which would interfere with diffraction due to the production of bubbles in the *operando* X-ray cell. The incident X-ray energy was set to the Co K-edge (E = 7722 eV), and the diffraction signal was recorded with an area detector (Pilatus 300K) positioned 1.5 m downstream from the *operando* cell at a rate of 1.9 mV per image (equivalent to 1.9 s per image). Figure 1b depicts the time-dependent intensity of the 211 tetragonal spinel peak and the background intensity at the Co K-edge (see the sketch in Figure 1a). Both intensities oscillate, and the period of both intensities matches the period of the applied potential during cyclic voltammetry. The intensity of the 211 Bragg peak varies as a function of applied voltage due to the evolving Co-oxidation state: the structure factor exhibits a prominent peak at resonance, and a shift in this resonance triggers a corresponding modification in intensity when measured at a fixed photon energy. The background intensity is due to X-ray fluorescence in the *operando* X-ray cell: the incident X-rays excite Co atoms in the catalyst, which relax and emit mutually incoherent fluorescent photons that produce broad background. Therefore, background intensity serves as a reporter for Co absorption, which changes when measured at a fixed energy as the absorption edge shifts in energy due to changes in Co oxidation state.[10]

We first analyze the Bragg reflections without considering the resonant nature of the structure factor by looking at the peak position, which is inversely proportional to the average lattice constant. The 211 diffraction peaks were fitted with a Gaussian function to determine the peak position, intensity, and full width at half maximum (FWHM). The average lattice spacings were calculated using Bragg's law[19] from the fitted peak positions and converted to strain. Strain was calculated as an engineering strain[20] using the percent change in the average lattice spacing at a certain voltage with respect to the average lattice spacing at 1.4 V and is shown in Figure 2. Peak width, which is affected by particle size and strain gradients, did not show a change that was significantly larger than the experimental uncertainty during cyclic voltammetry (Fig. S2). The intensity of the peak is the REXS signal (Fig. S3).

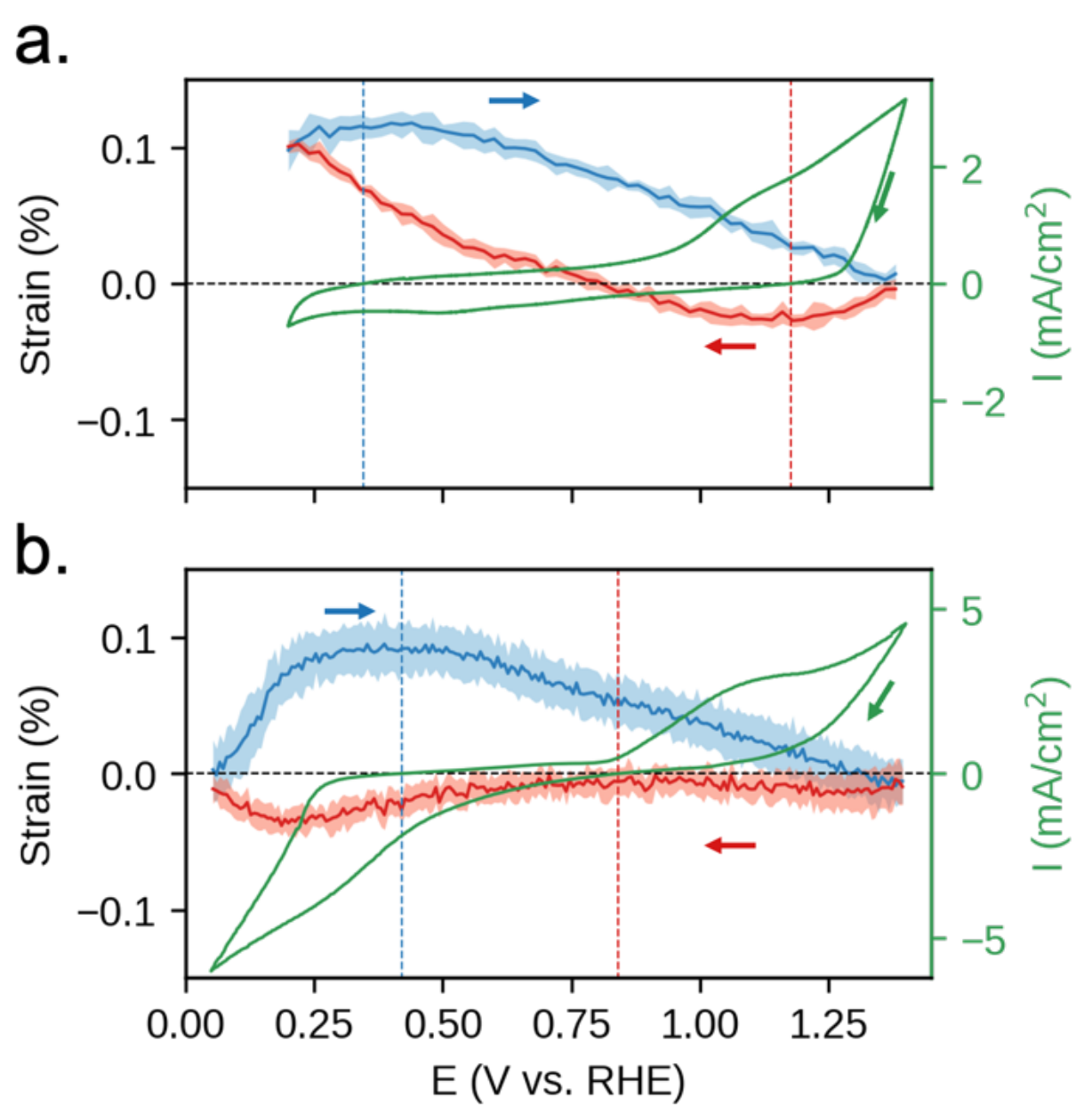

*Figure 2. Reversible strain changes of Co-Mn spinel oxides during cyclic voltammetry. (a.) Current during cyclic voltammetry (CV) from 0.2 to 1.4 V in green. Strain during cyclic voltammetry in red (negative sweep) and blue (positive sweep). (b.) Current during cyclic voltammetry from 0.05 to 1.4 V in green. Strain during cyclic voltammetry in red (negative sweep) and blue (positive sweep). The shaded areas in red and blue indicate the standard error in strain for all six cycles at each potential. Arrows indicate the direction of the potential sweep. The horizontal dashed lines in black indicate zero current and zero strain. The vertical dashed lines indicate potentials at which the current passes through zero; positive to negative in red and negative to positive in blue.*

Figure 2 shows strain in Co-Mn oxides as a function of applied potential during cyclic voltammetry. The solid lines in red and blue in Figure 2 represent strain averaged over the second through sixth cycles of cyclic voltammetry (Figure S1). We chose cycles 2-6 because of the visible decline of the diffraction signal after 20 cycles due to cell degradation in strongly alkaline media (Figure S1). The red curves in Figure 2 are strain during the negative potential sweeps, and the blue curves are strain during the positive sweeps. The green curves show cyclic voltammetry profiles. Figure 2a shows the strain and current behavior during cyclic voltammetry between 0.2 and 1.4 V vs. RHE. The strain values display hysteretic behavior with a local maximum in the blue curve and a local minimum in the red curve. The points of local maximum and local minimum strain occur when the current crosses zero (dashed vertical lines). This indicates that the direction of strain evolution changes when the current changes direction. The hysteresis between the forward and backward strain curves is likely due to the hysteresis in cyclic voltammetry from diffusion-limited processes through the electrolyte and at the electrolyte/catalyst interface.

Figure 2b shows the strain and current behavior during cyclic voltammetry between 0.05 and 1.4 V. The low potential cyclic voltammetry shows a nearly identical maximum strain of around 0.1% when compared to the strain behavior in Figure 2a. This maximum strain similarly occurs when the current switches from negative to positive in both Figure 2a and 2b. In fact, both potential ranges result in similar oxidative strain behaviors above 0.4 V during cyclic voltammetry (shown in blue). Differences emerge in the reductive potential sweep (shown in red), as strain remains nearly constant in Figure 2b. The larger potential window also results in a rapid lattice expansion when the potential is swept in a positive direction while the current is still negative, which corresponds to an increase in Co oxidation state (Fig. S3). Strain behaviors for both cyclic voltammetry conditions are reversible as shown by the small error bars, indicating robust structural reversibility and stability during normal fuel cell operating conditions. Nevertheless, the discrepancy between the red lines in Figures 2a and 2b indicates that the material exhibits distinct behavior at increased voltages after being subjected to 0.05 V, in contrast to when it is subjected to 0.2 V.

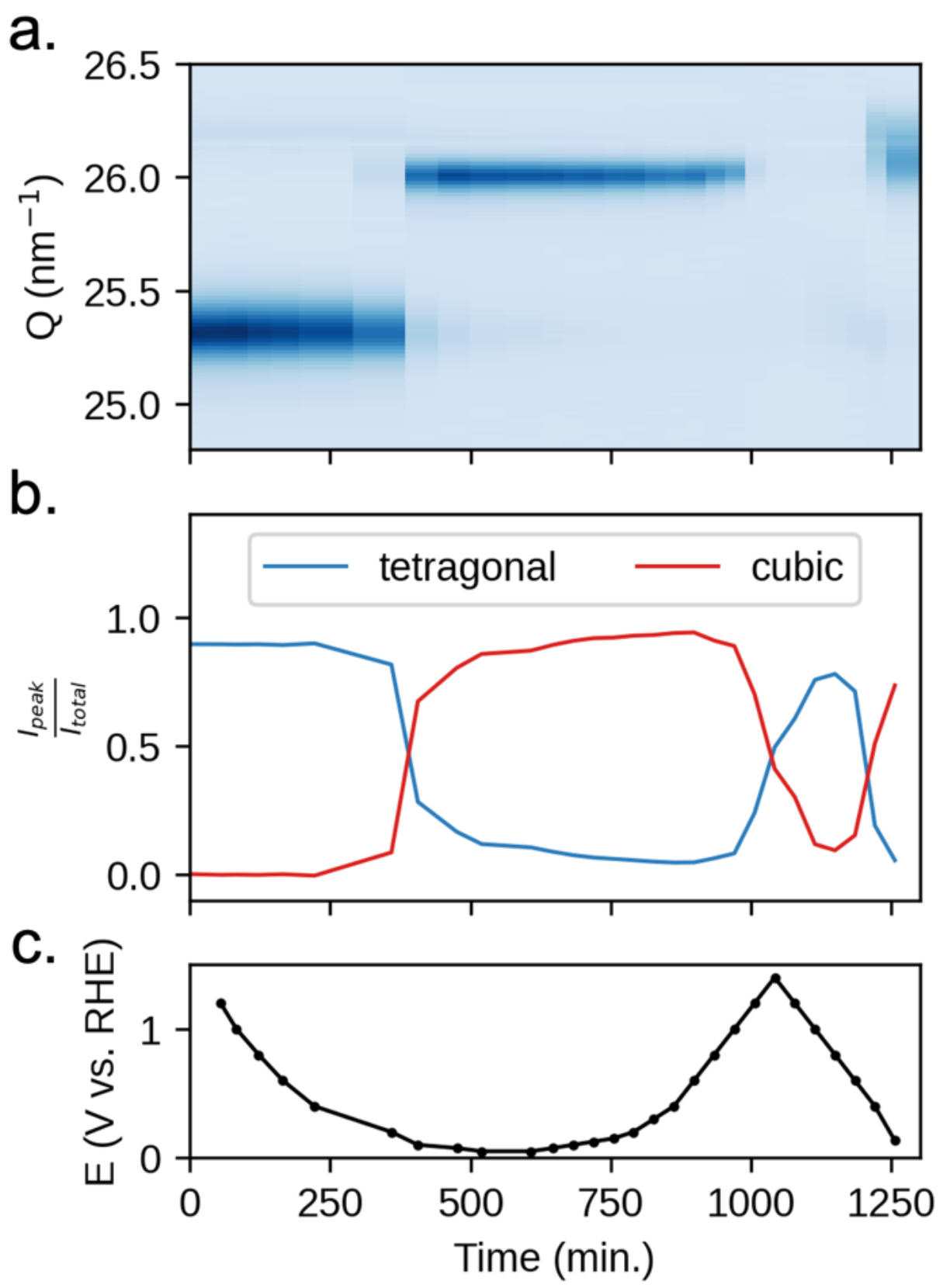

*Figure 3. Partially reversible phase transformation in $MnCo_2O_4$ under extreme ORR conditions. (a.) Operando XRD intensity during potentiostatic measurements. (b.) Peak intensity ratios for tetragonal (blue) and cubic (red) phases. (c.) Potentiostatic values for each scan.*

The Co-Mn spinel was also studied under *operando* conditions at constantly held voltages to gain further insights into the limit of structural stability during extreme ORR conditions. X-ray diffraction patterns were acquired at a series of incident X-ray energies around both the Co and Mn edges after the electrochemical current had decayed to a background level (i.e., a steady state). Figure 3a shows the 211 tetragonal spinel diffraction peak over a series of potentiostatic measurements ranging from 1.4 to 0.05 V. As the applied potential decreased, a structural phase change occurs below 0.2 V shown by the emergence of a new diffraction peak with a 2.5% decrease in layer spacing. This discontinuous change in layer spacing indicates a transition from a tetragonal spinel ($Mn_3O_4$ parent structure) to a cubic spinel ($Co_3O_4$ parent structure), with the new peak being the 311 peak of the cubic spinel. Figure S4 illustrates the difference between the extended c-axis of the tetragonal spinel compared to the cubic structure. This new 311 peak persists for the rest of the applied potentials but shrinks when returning to high potentials above 1 V where the 211 tetragonal peak faintly reappears (Fig. S5). Upon the second decrease in potential, the cubic 311 peak reemerges but with a significantly wider peak width, indicating partial reversibility with an increase in structural disorder. Figure 3b shows the intensity ratio of the tetragonal and cubic peaks in blue and red and illustrates the structural transition during the initial decrease in potential and the partial reversibility in the second potential cycle.

Additionally, *operando* REXS was used to probe the oxidation state of both Mn and Co during the tetragonal-cubic phase transformation. Figure 4a shows the REXS spectroscopy data, and Figure 4b shows the corresponding *operando* XRD peaks at 0.1 V, during two-phase co-existence. Here, the REXS spectrum of Mn cations incorporated in the emergent cubic phase (shown in red) is dominated by noise since the Mn REXS data acquisition preceded both the Co REXS and the diffraction (the latter was collected at an energy above the Co-edge) and is therefore probing slightly earlier in the transition. Nevertheless, the REXS spectrum of Mn cations incorporated in the tetragonal phase shows a minimum at 6.555 keV. At this same potential, two distinct Co REXS spectra are observed. The REXS spectrum of Co cations incorporated into the cubic structure (measured on the 311 peak) exhibits a minimum at 7.724 keV, while the REXS spectrum of Co cations in the tetragonal structure (measured on the 211 peak) shows a minimum 5 eV higher, highlighted by the red and blue vertical dashed lines in Figure 4a. This indicates that the Co cations in the emerging cubic phase are reduced compared to the ones in the tetragonal phase. Importantly, both Co spectra are measured simultaneously, which highlights the power of *operando* REXS by simultaneously distinguishing/probing the oxidation state of two different structural phases during a phase transition. Figure S6 compares the Co spectra at 0.1 V to those before and after the phase transition. Figures 4c, and d show *operando* diffraction and REXS spectroscopy data further along in the phase transition at 0.075 V. Here, diffraction indicates that the tetragonal phase fraction is decreasing, wthe cubic fraction is increasing. Figure 4c shows overlapping Mn spectra for the two phases revealing that both the tetragonal and cubic phases have a nearly identical Mn oxidation state. Nevertheless, it is important to note that these spectra are shifted to 5 eV lower photon energy when compared to the 0.1 V tetragonal Mn spectrum (dashed blue line in Figure 4c) indicating that both phases are reduced at 0.075 V compared to 0.1 V. A similar behavior is observed in the Co spectra. Both phases share Co oxidation state at 0.075 V, which is reduced compared to the 0.1 V tetragonal Co spectrum (dashed blue line). These observed reductive shifts are highlighted by the black arrows in Figure 4c.

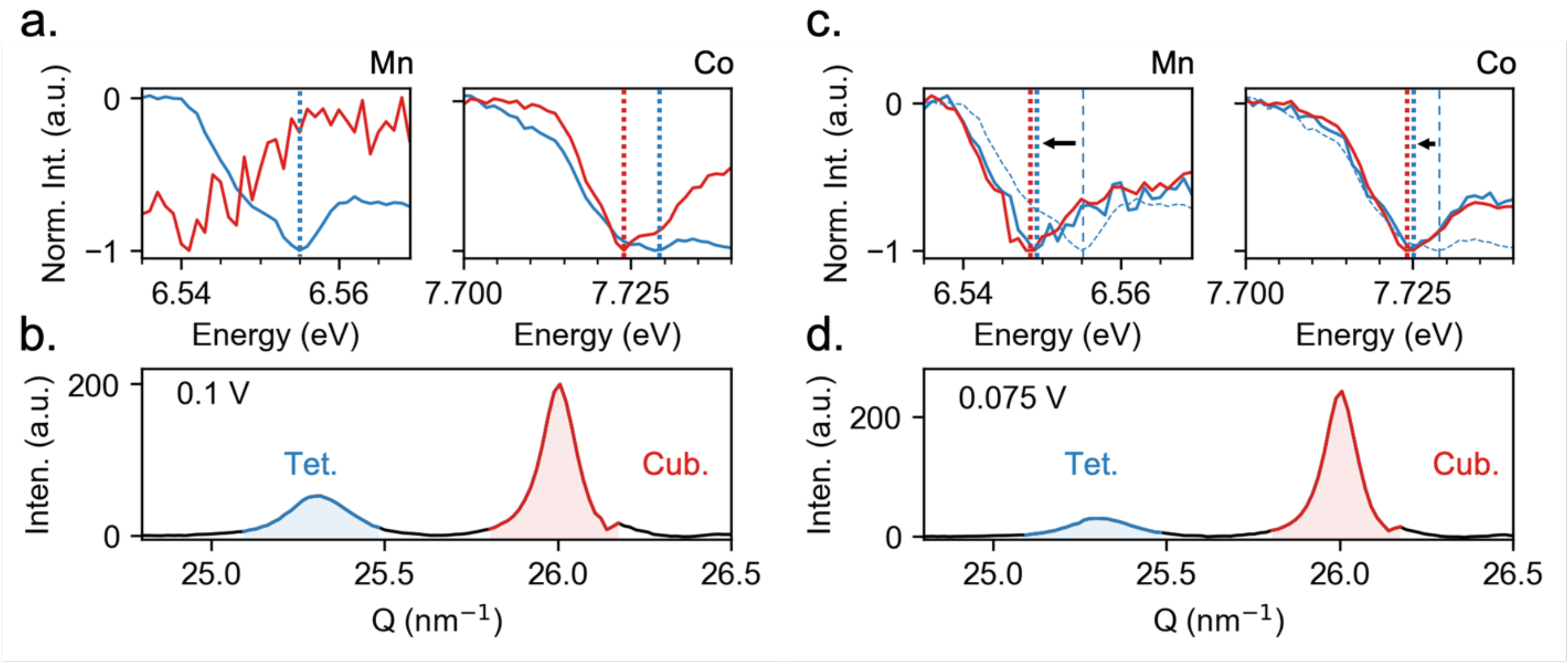

*Figure 4. Phase-specific spectroscopic changes during potentiostatic measurements. (a.) Co and Mn spectra for tetragonal (blue) and cubic (red) phases at 0.1V vs RHE. Vertical dashed lines indicate*

*spectrum minima. (b). Peak intensities for tetragonal (blue) and cubic (red) phases at 0.1V vs. RHE. (c.) Co and Mn spectra for tetragonal (blue) and cubic (red) phases at 0.075V vs. RHE. Tetragonal spectra at 0.1V shown with dashed blue lines. Vertical dashed lines indicate spectrum minima. Arrows highlight reductive shifts between 0.1V and 0.075V. (d). Peak intensities for tetragonal (blue) and cubic (red) phases at 0.075V vs. RHE.*

**Discussion**

To better understand the origins of the observed strain during CV, as shown in Figure 2, joint density-functional theory (JDFT) and stress-strain modeling were used to determine if adsorbed reactants during cyclic voltammetry could give rise to the strain observed with *operando* XRD. Previous JDFT work has shown the ability to accurately calculate the binding states of the 4-step oxygen reduction reaction on $Co_3O_4$ 001 surface.[21] A useful value during these calculations is the chemical surface stress tensor, which quantifies the stress state induced when a species is bound on the $Co_3O_4$ 001 surface. Even though $Co_3O_4$ and $MnCo_2O_4$ have different compositions, their mutual spinel structure and comparable properties make $Co_3O_4$ a convenient and effective model to inform our comprehension of Co-Mn oxide spinels.

To determine the surface strain induced by each individual adsorbed species, we used JDFT to compute the stress tensor per unit concentration of the 001 surface with a given adsorbed state for each of the 4 steps in the ORR, referred to as the chemical surface stress tensor. The surface strain induced by each individual adsorbed state can then be calculated via Hooke's law[22] by multiplying the compliance matrix of $Co_3O_4$[23,24] and the chemical surface stress tensor of each adsorbate state (*H+*OH, *H+*(O)(O), *H+*(OH)(O), *H+*(OH)(OH) ).[21] Moreover, since adsorbate coverage depends on voltage, the chemical surface stress tensor will change with shifts in voltage. Using JDFT adsorption energies,[21] we computed the adsorbate coverages as a function of applied potential and combined that with the corresponding chemical surface stress tensors to calculate the expected in-plane surface strain as a function of applied potential.

Stress-strain modeling is needed to convert the in-plane surface strain to the average strain throughout an entire particle, which is what *operando* XRD probes. We used a spherical thin-walled pressure vessel model[20] to estimate how the surface stress would propagate to the interior of a spherical nanoparticle. Specifically, we took the applied surface hoop stress of a spherical cross-section and distributed it across the entire cross-section (see Figure 5b). This results in the following equation relating the surface stress ($\sigma_s$) to the bulk stress ($\sigma_b$) as a function of the spherical radius (r) and thickness of the surface stress layer (t):[20]

$$\frac{\sigma_b}{\sigma_s} = \frac{2rt - t^2}{r^2}$$

Using this spherical thin-walled pressure vessel model and the linear relationship between stress and strain, average particle tensile strain is calculated for a 30 nm diameter nanoparticle with the surface stress layer of one lattice constant (0.8 nm) thick. Figure 5a shows the calculated values of average relative particle strain for potentials from 0.7 to 1.4 V vs. RHE. Below 0.7 V, the relative strain was calculated to be constant at around 0.018% tensile strain as all four reactant adsorbates are predicted to be present in equal proportions. This value for the particle tensile strain is significantly smaller than what was observed in the *operando* XRD data. The discrepancy suggests that surface adsorbed reactants are not the main cause of the strain behavior seen during cyclic voltammetry and that reactants might be diffusing into the interior of the nanoparticles via pores or intercalation.[8,25]

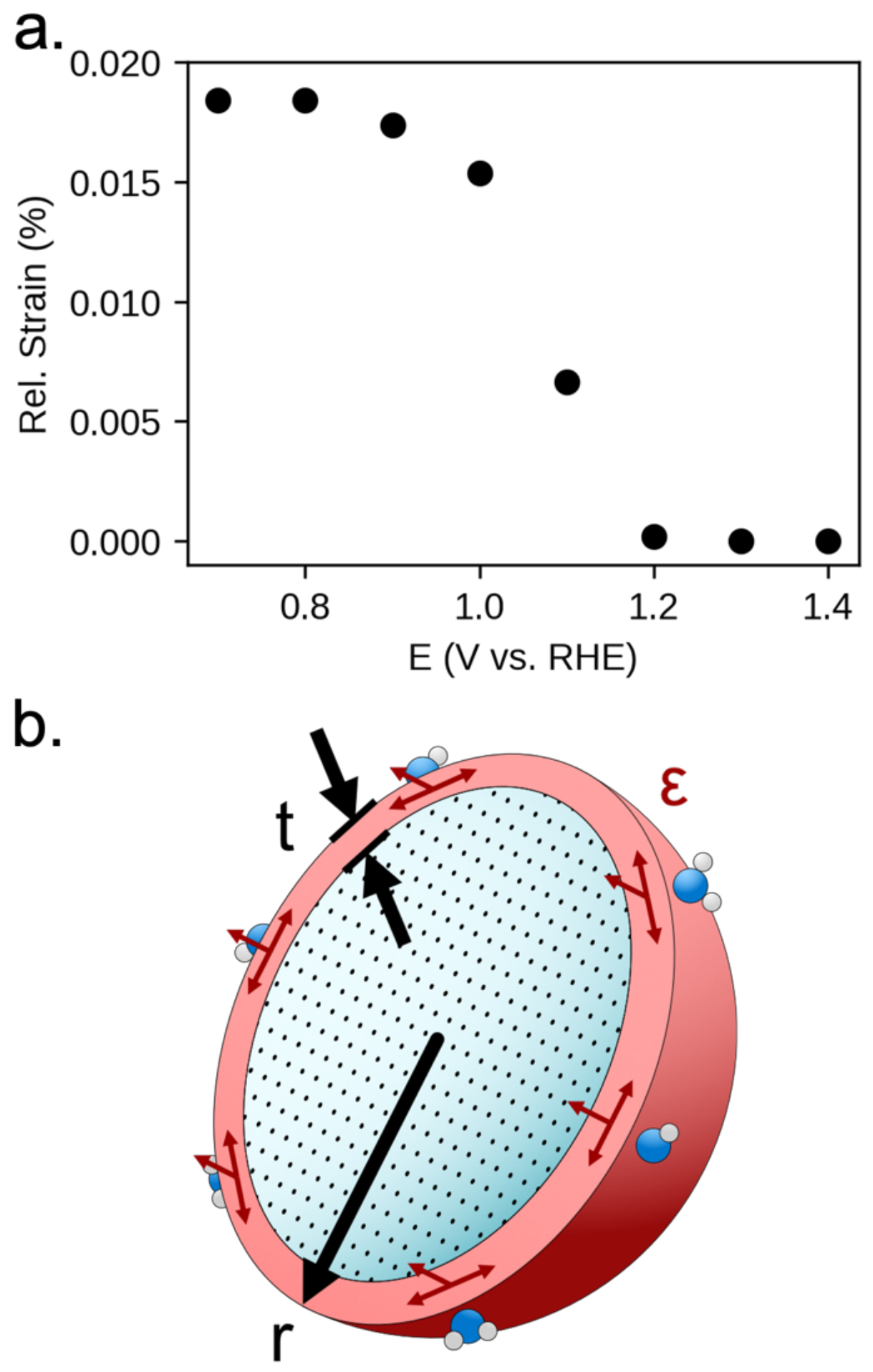

*Figure 5. (a.) Calculated average relative particle strain as a function of applied potential. (b.) Illustration of thin-walled pressure vessel modeled nanoparticle with adsorbed species. In-plane biaxial stress can induce shrinking or swelling of the particle.*

While JDFT and strain modeling calculations indicate that surface adsorption during the ORR is not large enough to be the main cause of the positive reversible strain observed during short low-potential exposure (cyclic voltammetry), extended low-potential exposure (constant potential) exhibits an irreversible phase transition. This suggests that there is a time-dependent process occurring at low potentials, which is kinetically limited during cyclic voltammetry. One possible reason for this behavior is intercalation. Intercalants such as the ORR reactants (e.g., *H, *O, *OH)[21] could diffuse into the interior of the nanoparticles. Previous studies have shown that water intercalation in $Mn_3O_4$ occurs as $Mn_2O_4$ forms via Mn dissolution.[25] As intercalants diffuse into the spinel and occupy interstitials, a resulting increase in lattice constant is possible. This intercalation process has also been shown to induce structural phase transformations,[25] which might serve as an explanation for the tetragonal to cubic spinel transformation.

From previous studies of spinel oxides, it has been revealed that the tetragonal symmetry of $Mn_3O_4$ is induced by Jahn-Teller distortions on $Mn^{3+}$.[26] The observed reduction of Mn is a possible driving force for the observed phase transition. As $Mn^{3+}$ is reduced to $Mn^{2+}$ and all d orbitals are partially filled, it loses degeneracy and Jahn-Teller character. Therefore, cubic spinel symmetry is preferred. As suggested by previous studies of Co-Mn oxide spinels, Mn reduction is possibly driven by oxygen vacancy formation.[8] Yet, the existence of reduced Mn in a tetragonal phase (see Figure 4c) may indicate a possible metastable state where the surrounding crystalline structure necessitates portions of the nanoparticle to stay in the tetragonal symmetry despite the cubic structure being preferred by the Mn oxidation state. Further studies are needed to confirm these hypotheses for the origin of the strain and oxidation state behavior observed by *operando* XRD and REXS.

## Conclusion

In this work, a mechanistic understanding of the synergistic Co-Mn ORR is revealed using *operando* XRD and REXS. *Operando* XRD of $MnCo_2O_4$ nanoparticles revealed significant strain induced during dynamic potential conditions, indicating structural robustness during operating conditions. Under constant potential conditions at 0.05 V vs. RHE, a phase transformation is observed accompanied by a reduction of both Co and Mn, which provide information/insights into the degradation mechanisms in Co-Mn spinel electrocatalysts at extremely low voltages. While it is unclear what the origins of the strain and phase transformation are, JDFT calculations reveal that adsorbing reactants are not the major cause of strain in Co-Mn oxide spinels during the ORR at the particle surface. Intercalation and oxygen vacancy formation are proposed as mechanisms that could be the origin of phenomena observed during this multimodal *operando* X-ray study based on the reduction of Co and Mn and previous studies. *Operando* XRD and REXS provide unique dynamic information that deepens the understanding of the complex interactions impacting the activity and long-term stability of Co-Mn spinel oxide electrocatalysts.

# Acknowledgments

Research primarily supported as part of the Center for Alkaline Based Energy Solutions (CABES), an Energy Frontier Research Center funded by the U.S. Department of Energy (DOE), Office of Science, Basic Energy Sciences (BES), under Award # DE-SC0019445. This work is based on research conducted at the Center for High-Energy X-ray Sciences (CHEXS), which is supported by the National Science Foundation (BIO, ENG and MPS Directorates) under award DMR-1829070.

# Supporting Information

**Synthesis of $MnCo_2O_4/C$**: Detailed synthesis methods can be found in our previous study.[1] Briefly, $MnCo_2O_4/C$ was synthesized using a facial hydrothermal method. Manganese (II) acetate was dissolved in 15 mL deionized (DI) water and sonicated for 15 min. 500 μL of concentrated $NH_3{\cdot}H_2O$ were diluted in 5 mL of DI water and added to the metal precursor solution dropwise under vigorous stirring at 1200 rpm. The pH of the formed metal-$NH_3$ complex solution was tested to be around 11. Ethanol (20 mL) was later added to the metal-$NH_3$ complex solution with an EtOH/$H_2O$ volume ratio of 1:1. High- surface-area carbon Ketjen Black (HSC KB) was weighted to achieve target metal oxide loadings of 40 wt.% in the catalyst/carbon composites. HSC KB was added to the suspension solution and stirred at 1200 rpm and 60 °C for a 12 h ageing process. The solution was then transferred into a 50 mL autoclave for hydrothermal reaction at 150 °C with a pressure of 30 bar for 3h. $MnCo_2O_4$ and NPs supported on carbon was separated from the residual solution using a centrifuge at 6000 rpm and washed with EtOH/$H_2O$ (vol. 1:1) three times and dried in oven at 80 °C for 6 h. Powder X-ray diffraction (XRD) patterns were collected at a scan rate of 2 °/min at 0.02° steps from 20° to 80° using a Rigaku Ultima IV Diffractometer.

**Operando XRD and REXS:** The experiment was conducted at the QM2 beamline at the Cornell High Energy Synchrotron Source (Cornell University, Ithaca, NY). The operando electrochemical cell window was aligned to the X-ray beam. Cyclic voltammetry was conducted at a sweep rate of 1mV/s and potential windows of 0.2 to 1.4 V vs. RHE and 0.05 to 1.4 V vs. RHE. For cyclic voltammetry with lower potential limit of 0.05 V and potentiostatic measurements, a Pilatus 300K detector was placed 1.5 m downstream from the electrochemical cell to measure the 211 tetragonal spinel and 311 cubic spinel peaks. For cyclic voltammetry with a lower potential limit of 0.2, a Pilatus 100K detector was placed 0.5 m downstream from the electrochemical cell. During cyclic voltammetry to 0.05 V measurements, the photon energy was set to 7722 eV, and diffraction images were collected at a rate of 1.9 s per image. During cyclic voltammetry to 0.2V measurements, the photon energy was set to 7720 eV, and diffraction images were collected at a rate of 14 s per image. During potentiostatic measurements, the photon energy was scanned repeatedly from 7700 to 7740 eV in 36 points at 1 second per energy for Co REXS and from 6535 to 6569 eV in 35 points at 2 seconds per energy for Mn REXS.

**X-ray Data Processing:** 2D diffraction images taken by the Pilatus detectors were processed by 1D azimuthal integration using PyFAI[2] to generate 1D diffraction patterns. To calculate background intensity, a quartic function was fit to the background signal of the integrated diffraction pattern. To calculate peak intensity, width, and position, background subtracted diffraction patterns were fit with a Gaussian function. Additionally, PyMCA[3] was used to extract experimental information generated by beamline QM2 at the Cornell High Energy Synchrotron Source (CHESS).

**Electrochemical Experiments:** 20 mg of $MnCo_2O_4$/C (40% mass loading) were dispersed in 50 mL of a Nafion (0.05% mass fraction) ethanol solution. Carbon paper (200 µm thick) was used as the substrate for the catalyst layer and the skeleton of the carbon paper was filled with carbon powder to increase the surface area and its mechanical strength for later device assembly. Carbon paper was later tailored into $1 \times 5$ $cm^2$ pieces as catalyst support. The catalyst-ionomer mixture was sprayed on one end of the carbon paper ($1 \times 1$ $cm^2$) using an airbrush and the rest, $1 \times 4$ $cm^2$, served as a non-active conductor with negligible effects on the catalytic current, compared with the active metal oxide catalysts.

The electrochemical cell includes two pieces of PEEK, which is chemically inert in the strong alkaline conditions of 1M KOH. The two PEEK pieces could be tightly sealed using six screws. A Teflon U-shaped sealing ring was placed between the two PEEK pieces to seal the electrolyte into the cell. On top of the electrochemical cell, two holes were used as a gas inlet and outlet. Inside the electrochemical cell, the section of the carbon paper with the catalyst layer was immersed into the electrolyte near the window for X-ray transmission. A carbon rod was used as the counter electrode (CE) and placed near the working electrode (WE). Ag/AgCl (sat. KCl) was used as the reference electrode (RE) and was connected to the cell via a thin plastic tube containing the electrolyte, which served as a salt bridge. 1M KOH was used as the electrolyte and saturated with $N_2$ gas to exclude dissolved $O_2$.

Cyclic voltammetry was conducted at 1mV/s, and two potential windows were used, 0.05 to 1.4V vs. RHE and 0.2 to 1.4V vs RHE. Potentiostatic measurements were done by holding potential and allowing the resulting current to decay to background levels before X-ray images were taken. The following potentials were used: 1.20, 1.00, 0.800, 0.600, 0.400, 0.200, 0.133, 0.100, 0.075, 0.050 V vs. RHE.

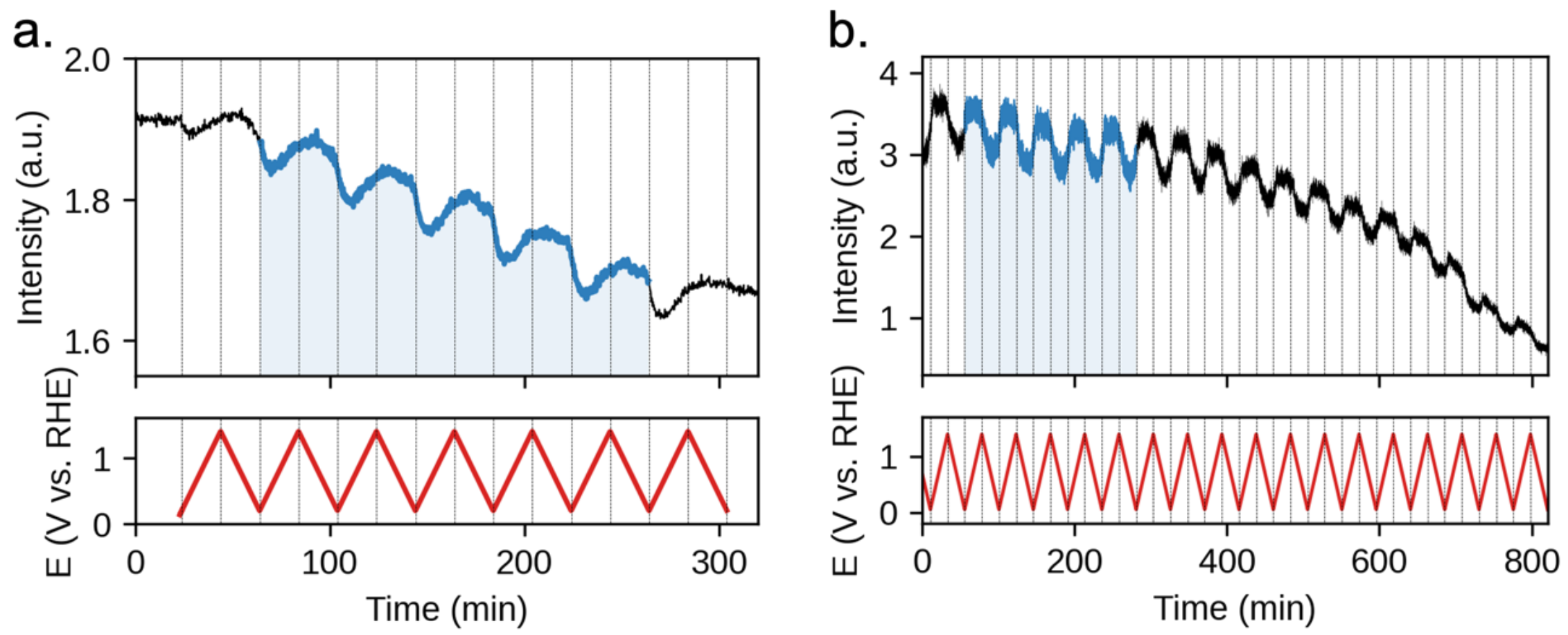

Figure S1. *Operando* REXS intensity and applied voltage over time during cyclic voltammetry. (a.) REXS intensity during CV from 0.2 to 1.4V vs. RHE. (b.) REXS intensity during CV from 0.05 to 1.4V vs. RHE. Blue highlighted area indicates data used for Figure 2, S1, and S2.

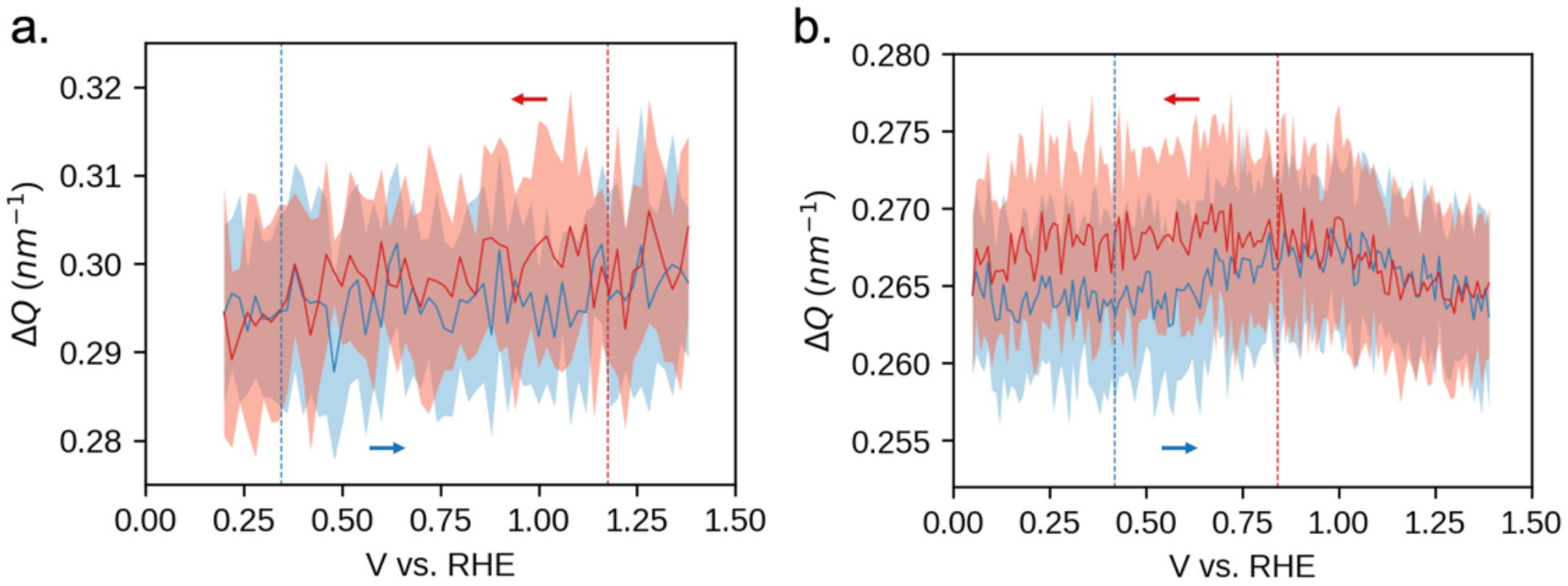

Figure S2. *Operando* XRD peak width (FWHM) during cyclic voltammetry. (a.) Peak width during CV from 0.2 to 1.4V vs. RHE. (b.) Peak width during CV from 0.05 to 1.4V vs. RHE. The red lines show peak width during negative potential sweep indicated with red arrow. The blue lines show peak width during positive potential sweep indicated with blue arrow. Shaded areas indicate standard error of the five cycles averaged. The vertical dashed lines indicate potentials at which the current passes through zero; positive to negative in red and negative to positive in blue.

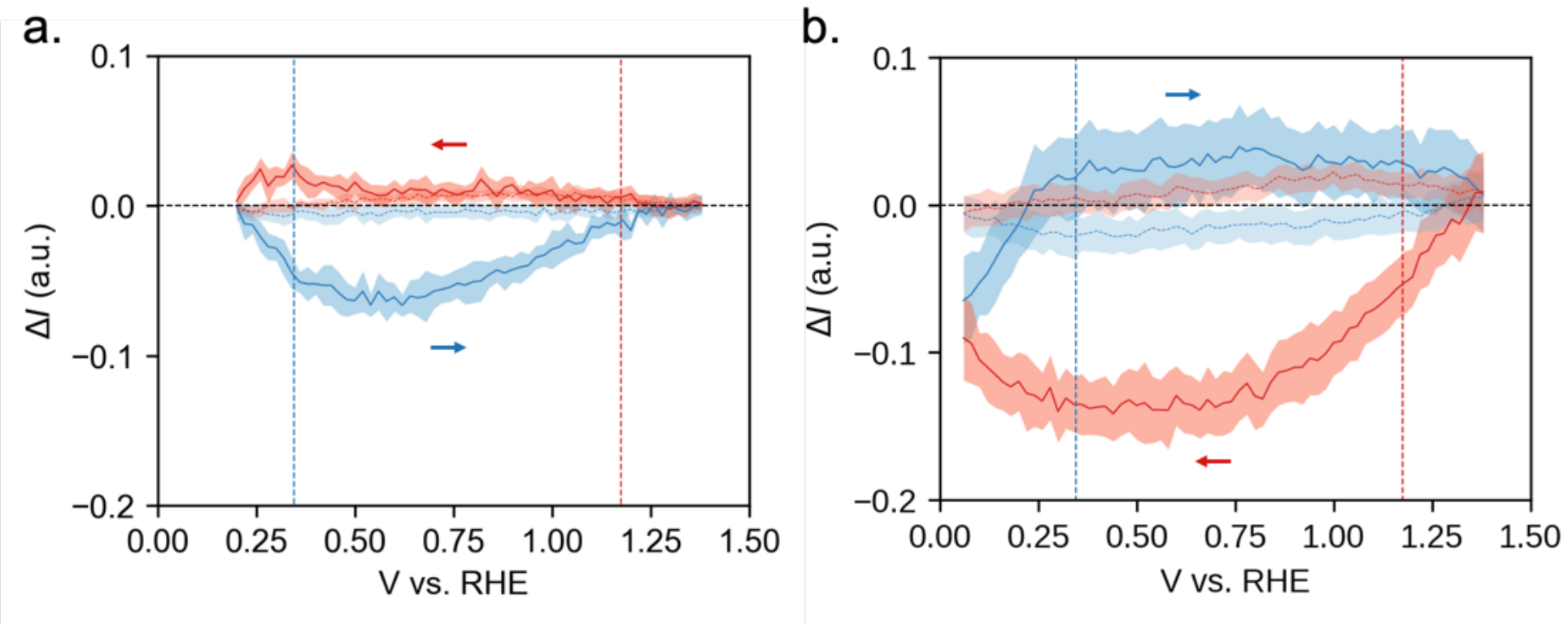

Figure S3. *Operando* REXS change in peak intensity and background fluorescence intensity during cyclic voltammetry (CV) compared to intensity at 1.4V. (a.) Peak and background intensity during CV from 0.2 to 1.4V vs. RHE. (b.) Peak and background intensity during CV from 0.05 to 1.4V vs. RHE. Solid lines indicate peak intensity while dashed lines indicate background fluorescence intensity averaged over five cycles. The red lines show peak width during negative potential sweep indicated with red arrow. The blue lines show peak width during positive potential sweep indicated with blue arrow. Shaded areas indicate standard error of the five cycles averaged. The horizontal dashed line in black indicates zero change in intensity. The vertical dashed lines indicate potentials at which the current passes through zero; positive to negative in red and negative to positive in blue.

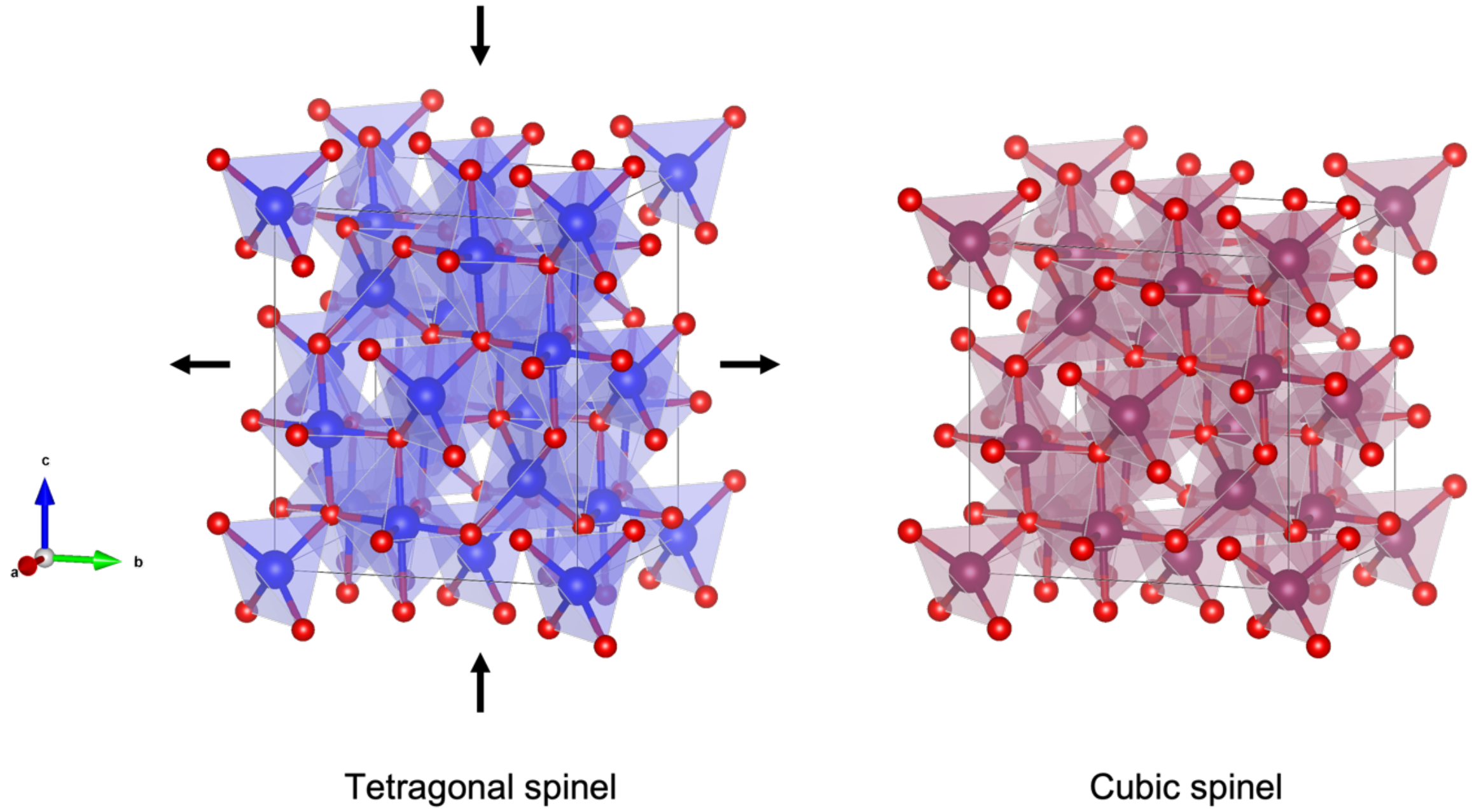

Figure S4. VESTA[4] visualization of tetragonal and cubic spinel structures. Oxygens are depicted in red and transition metals (Co/Mn) are depicted in blue (tetragonal) and maroon (cubic). Arrows indicate shortening of the c-axis and expansion of the a and b-axes during the transition from tetragonal to cubic spinel.

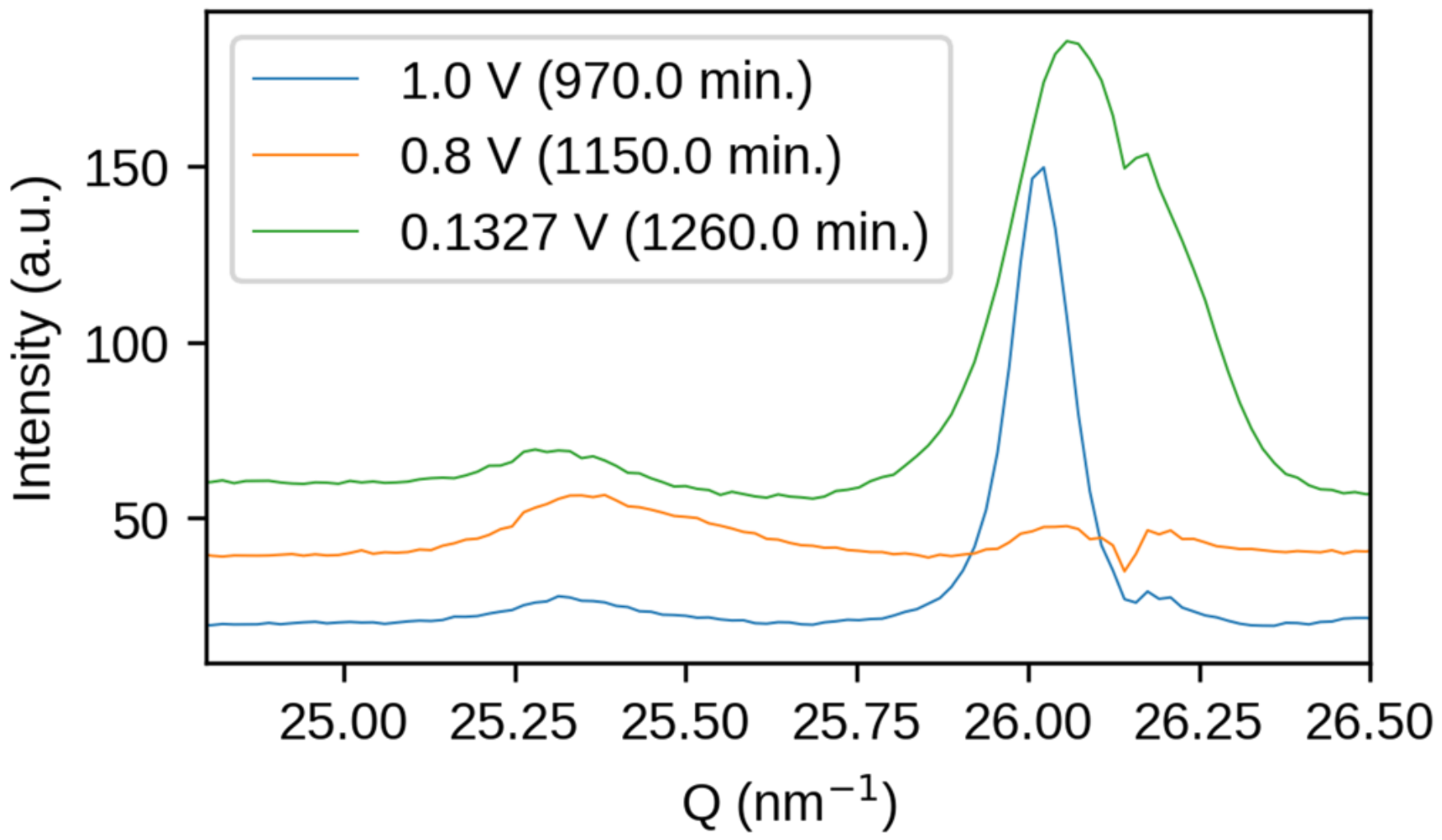

Figure S5. *Operando* XRD intensity during the second potentiostatic cycle. At 0.8V during the second cycle (orange), there is a significant decrease in the cubic peak intensity and a slight increase in the tetragonal peak intensity. At lower voltages in the second cycle (green), the cubic peak intensity increases significantly and is wider than in the first cycle (blue). Diffraction intensities are shifted for visibility.

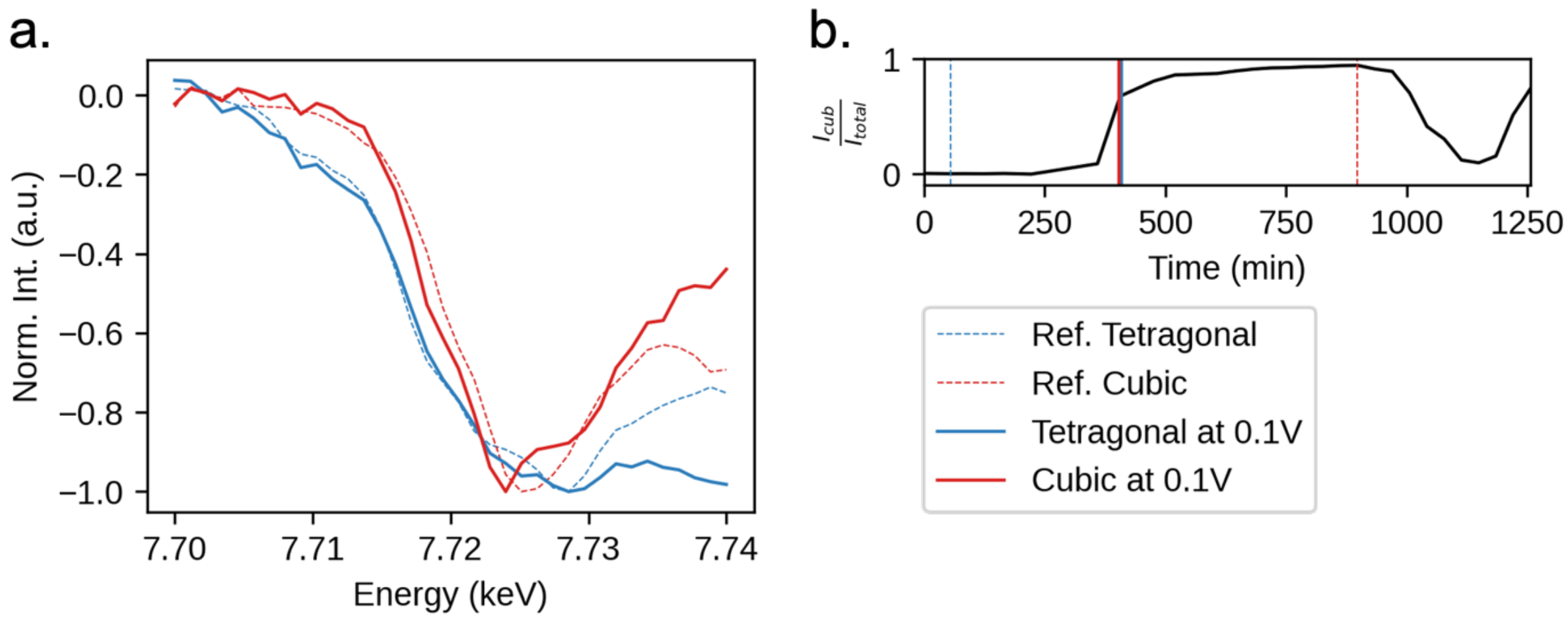

Figure S6. REXS Co spectra before, during, and after phase transition. Dashed lines show Co REXS spectra before, tetragonal phase (blue), and after, cubic phase (red), the structural phase transition. Solid lines show the REXS spectra of the tetragonal (blue) and cubic (red) phases at 0.1V during the phase transition.

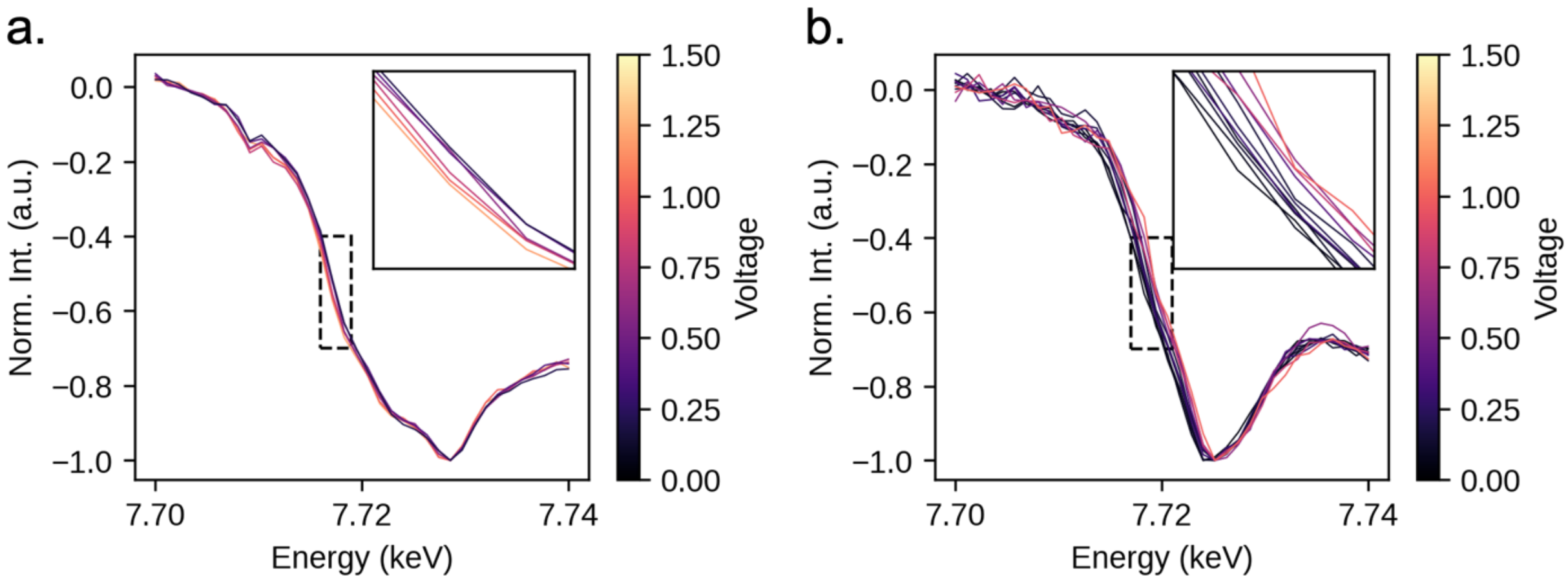

Figure S7. REXS Co spectra. (a.) Tetragonal spinel Co REXS spectra during potentiostatic voltage decrease. (b.) Cubic spinel Co REXS spectra during potentiostatic voltage increase.

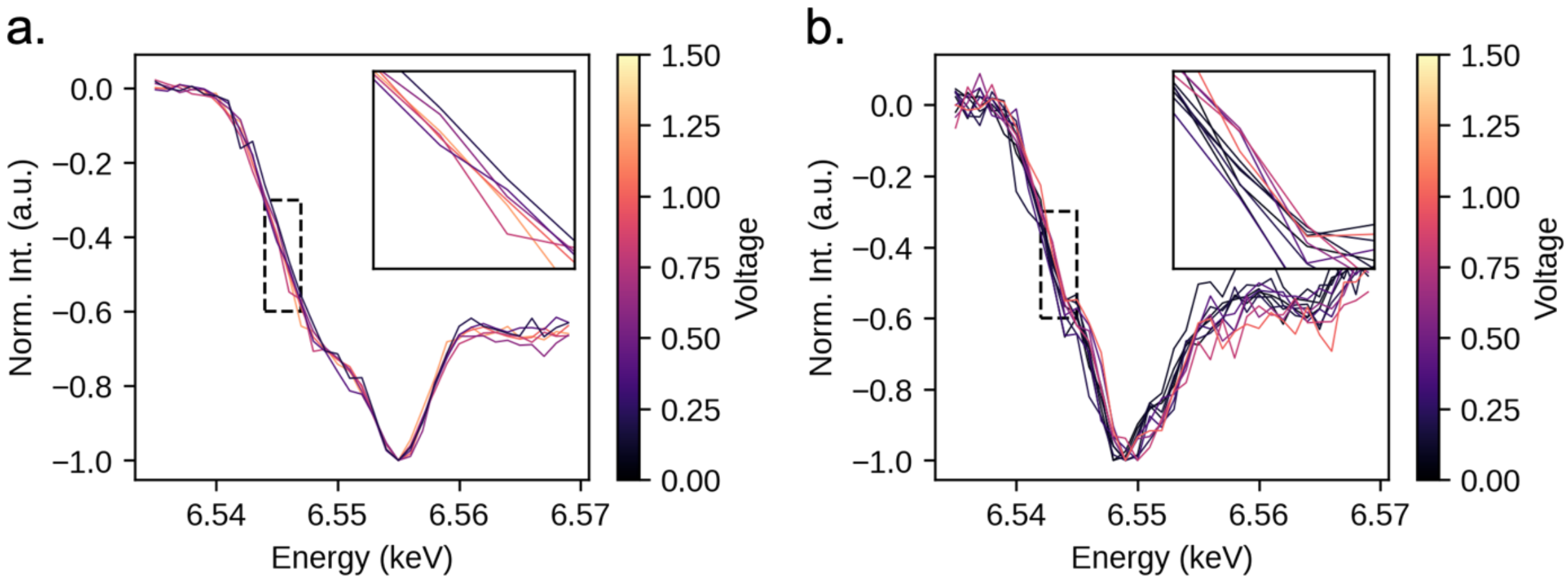

Figure S8. REXS Mn spectra. (a.) Tetragonal spinel Mn REXS spectra during potentiostatic voltage decrease. (b.) Cubic spinel Mn REXS spectra during potentiostatic voltage increase.

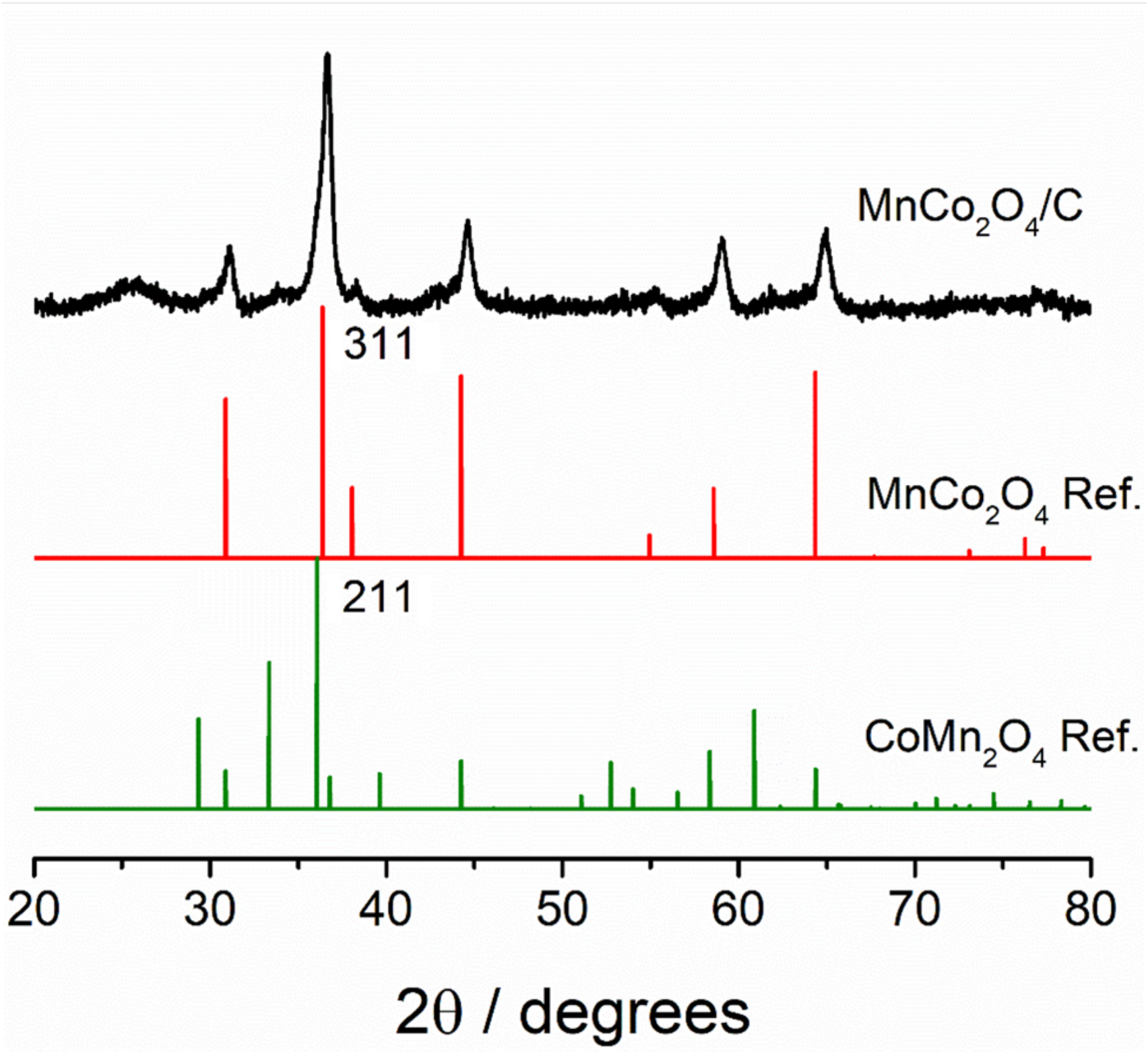

Figure S9. Powder XRD patterns of $MnCo_2O_4$/C with reference patterns of cubic $MnCo_2O_4$ (PDF# 01-084-0482) and tetragonal $CoMn_2O_4$ (PDF# 01-077-0471).

# Supporting Information References